\documentclass[reprint, showpacs]{revtex4-1}
\usepackage{graphicx}
\usepackage{amsfonts}
\usepackage{natbib}
\usepackage{float}

\newcommand{\be}{\begin{equation}}
\newcommand{\ee}{\end{equation}}
\newcommand{\bea}{\begin{eqnarray}}
\newcommand{\eea}{\end{eqnarray}}

\newcommand{\al}{\alpha}
\newcommand{\bet}{\beta}
\newcommand{\la}{\lambda}
\newcommand{\mA}{\left< A \right>}

\begin{document}
\title{Statistical properties of the localization measure of
  chaotic eigenstates and the spectral statistics in a mixed-type billiard}

\author{Benjamin Batisti\'c}
\author{\v Crt Lozej}
\author{Marko Robnik}

\affiliation{CAMTP - Center for Applied Mathematics and Theoretical
Physics, University of Maribor, Mladinska 3, SI-2000 Maribor, Slovenia, European Union}

\date{\today}

\begin{abstract}
  We study the quantum localization in the chaotic eigenstates of a
  billiard with mixed-type phase space
  (J. Phys. A: Math. Gen. {\bf 16}, 3971 (1983); {\bf 17}, 1049 (1984)),
  after separating the regular and chaotic eigenstates,
  in the regime of slightly distorted circle billiard where
  the classical transport time in the momentum space is still large enough,
  although the diffusion is not normal. This is a continuation of our recent papers
  (Phys. Rev. E {\bf 88}, 052913 (2013); {\bf 98}, 022220 (2018)).
  In quantum systems with discrete energy spectrum
  the Heisenberg time $t_H =2\pi \hbar/\Delta E$, where $\Delta E$ is the mean level spacing (inverse
  energy level density), is an important time scale. The classical transport time scale $t_T$
  (transport time) in relation to the Heisenberg time scale $t_H$ (their ratio is the parameter
  $\alpha=t_H/t_T$) determines the degree of localization of the chaotic eigenstates,
  whose measure $A$ is based on the information entropy. We show that $A$ is
  linearly related  to normalized inverse participation ratio.
  The localization of chaotic eigenstates is reflected also in the
  fractional power-law repulsion between the nearest energy levels in the
  sense that the probability density (level spacing distribution)
  to find successive levels on a distance $S$ goes like $\propto S^\beta$ for
  small $S$, where $0\leq\beta\leq1$, and $\beta=1$ corresponds to completely
  extended states. We show that the level repulsion exponent $\beta$ is 
  empirically a rational   function of $\al$, and the mean
  $\langle A \rangle$ (averaged over more than 1000 eigenstates) as a function of $\al$ is
  also well approximated by a rational function.
  In both cases there is some scattering of the empirical data around the mean curve,
  which is due to the fact that $A$ actually has a distribution,
  typically with quite complex structure, but in the limit $\al\rightarrow \infty$
  well described by the beta distribution. The scattering
  is significantly stronger than (but similar as) in the stadium billiard
  (Nonl.Phen.Compl.Sys. {\bf 21}, No3, 225 (2018)) and the kicked rotator
  (Phys. Rev. E {\bf 91}, 042904 (2015)). Like in other systems,
  $\beta$  goes from $0$ to $1$ when $\al$ goes from $0$ to $\infty$.
  $\beta$ is a function of $\langle A \rangle$, similar to
  the quantum kicked rotator and the stadium billiard. 
\end{abstract}

\pacs{01.55.+b, 02.50.Cw, 02.60.Cb, 05.45.Pq, 05.45.Mt}

\maketitle

\section{Introduction}
\label{sec1}

Quantum chaos (or more generally, wave chaos) deals with phenomena in the quantum domain,
which are signatures of the classical chaos in the corresponding classical systems.
\cite{Stoe, Haake, Rob2016}. The classical dynamics as the ray dynamics of the quantum
wave functions is, for example, an analogy of the relationship between
the Gaussian ray optics and the wave phenomena of the Maxwell equations describing
the electromagnetic field. The classical and the quantum descriptions are connected
theoretically through the semiclassical mechanics, which is the short wavelength
approximation of the underlying wave field.

In the classically integrable Hamiltonian systems with $f$ degrees of freedom
the semiclassical theory predicts that the quantum eigenstates (in the $2f$-dim phase space,
represented by Wigner functions \cite{Wig1932} or Husimi functions \cite{Hus1940})
are associated with the
classical $f$-dim invariant tori. In the case of classical ergodic systems, the
eigenstates are microcanonical (uniformly spread over the $(2f-1)$-dim energy surface).
In the case of the mixed-type classical phase space, where regular regions covered by
the invariant tori coexist with the chaotic sea (one or more chaotic invariant regions),
we have the generic structure (almost all systems are of this type), where we have to
distinguish between the quantum regular and irregular (chaotic) eigenstates, an idea proposed
qualitatively already in 1973 by Percival \cite{Percival1973}. This line of thought lead to the
Principle of Uniform Semiclassical Condensation of Wigner functions \cite{Rob1998},
based on work of Berry \cite{Berry1977}, Shnirelman \cite{Shnirelman1974}, Voros \cite{Voros1979},
and further developed by Veble, Robnik and Liu \cite{VRL1999}. The Wigner functions
of the eigenstates condense uniformly on the classical invariant component in the
classical phase space, and this principle (PUSC) has a great predictive power as
demonstrated e.g. in Ref. \cite{VRL1999}.

The classical dynamics of bounded Hamiltonian systems determines also the statistical
properties of the discrete energy spectra of the corresponding eigenstates.
For the classically regular motion it predicts Poissonian energy level statistics, while
in the classically fully chaotic (ergodic) systems the statistics of Random Matrix Theory (RMT)
applies, as conjectured by Bohigas, Giannoni and Schmit \cite{BGS1984} in 1984, also by Casati,
Valz-Gris and Guarneri \cite{Cas1980},  proven by
Berry \cite{Berry1985}, Sieber and Richter \cite{Sieber}, and by Haake and coworkers
\cite{Mueller1,Mueller2,Mueller3,Mueller4},
using the semiclassical techniques based on the Gutzwiller's periodic orbit theory
(See Refs. \cite{Gutzwiller1967,Gutzwiller1969,Gutzwiller1970,Gutzwiller1971,Gutzwiller1980}
and also the books by St\"ockmann
\cite{Stoe} and Haake \cite{Haake}).

The intermediate case of the mixed-type Hamilton systems was treated first theoretically
by Berry and Robnik \cite{BerRob1984}, and has been  analyzed later on in many studies, most
accurately by Prosen and Robnik \cite{ProRob1999}. In this picture the parameter $\rho_1$
plays the crucial role, being the relative fraction of the phase space volume occupied by the regular
regions in the classical phase space, and it also is the relative density of the regular energy
levels in the total quantum spectrum of the underlying system. The spectral statistics
for the regular levels is Poissonian.

If there are chaotic regions with the relative volume fractions (and corresponding
energy level densities) $\rho_2, \rho_3, \dots$, then for each of them the RMT statistics
applies. Usually, the dominant chaotic region is by far the largest one, $\rho_2 \gg \rho_3,\dots$,
so that the smaller chaotic regions can be neglected, and we have $\rho_1+\rho_2 =1$.

The best mathematical description of such a mixed-type case is in terms of the gap
probability $E(S)$.  This is the probability that an energy interval (after unfolding,
i.e. reducing the mean energy level density to unity) is empty of levels. Clearly,
if regular and chaotic eigenstates are not correlated, being statistically independent
of each other, the gap probabiliy simply factorizes, that is

\be \label{gapprob}
E(S) = E_{P}(\rho_1S) \; E_{RMT}(\rho_2S),
\ee
where $P$ and $RMT$ refer to the Poissonian and
RMT statistics, respectively. The Poissonian is $E_P(S)=\exp (-S)$.
For the GOE level spacing distribution,
which applies if the time reversal symmetry (or any other antiunitary
symmetry) exists,
the well known Wigner distribution (Wigner surmise) is an excellent
analytical approximation

\be \label{Wignerdistr}
P_W(S) = \frac{\pi S}{2} \exp \left(-\frac{\pi S^2}{4} \right),
\ee
while the corresponding gap probability is

\be \label{WGP}
E_W(S) =  1 - {\rm erf} \left( \frac{\sqrt{\pi}S}{2} \right) 
 = {\rm erfc} \left( \frac{\sqrt{\pi}S}{2} \right).
\ee
The level spacing distribution $P(S)$ is the second derivative
of the gap probability $P(S)=d^2E(S)/dS^2$, and therefore in this case given by

\bea \label{BerryRobnik}
P_{BR}(S)  &=&   e^{-\rho_1 S}  e^{- \frac{\pi \rho_2^2 
S^2}{4}} \left( 2 \rho_1 \mu_2 + \frac{\pi \rho_2^3 S}{2} \right) \\ \nonumber
  &+&   e^{-\rho_1 S} \rho_1^2 {\rm erfc}
\left( \frac{\sqrt{\pi} \rho_2 S}{2} \right). 
\eea
as derived by Berry and Robnik \cite{BerRob1984}. Of course, $\rho_2=1-\rho_1$.
The gap probability $E(L)$ is just a special case of $E(k,L)$ probability
of finding $k$ levels on an interval of length $L$, namely $E(L)=E(0,L)$.
For more details about the $E(k,L)$ probabilities with $k>0$ see Ref. \cite{ProRob1999}. 

The above statements are correct only if the chaotic states are uniformly extended over the
classical invariant chaotic component. This condition, however, is not always satisfied.
The phenomenon of dynamical (or quantum) localization can occur, first discovered and
further explored by Chirikov, Izrailev and Shepelyansky \cite{Chi1981} in the
quantum kicked rotator (QKR) introduced by Casati, Chirikov, Izrailev and Ford \cite{Cas1979}
as a model system, and later extensively studied in particular by Izrailev
\cite{Izr1986,Izr1987,Izr1988,Izr1989,Izr1990}. The QKR is a time periodic (Floquet) system.
The time independent chaotic systems are exemplified by the 2-dim billiard systems.
Borgonovi, Casati and Li \cite{BCL1996} have studied from this point of view 
the stadium billiard of Bunimovich \cite{Bun1979}.
See also the review by Prosen \cite{Pro2000}. The case of mixed-type billiard
has been studied recently by Batisti\'c and Robnik \cite{BatRob2010,BatRob2013A,BatRob2013B}.

The criterion for localization is in terms of the ratio

\be  \label{alpha}
\alpha= \frac{t_H}{t_T}
\ee
of the Heisenberg time $t_H$ and the classical transport time $t_T$.
Here $t_H=2\pi\hbar/\Delta E$, with $\Delta E$ being the mean energy level spacing
(inverse energy level density), which is an important time scale in any
quantum system with discrete energy spectrum, while $t_T$ is the
purely classical ($\hbar$-independent)  diffusion time,
or typical time needed for an ensemble of initial sharply distributed
momenta to spread uniformly over the classical chaotic component.
If $\alpha \ll 1$ the chaotic eigenstates are maximally localized,
while if $\alpha \gg 1$ the eigenstates are maximally extended, but in
between we have the partially localized eigenstates. The degree of localization
can be measured most easily in terms of the Husimi function \cite{Hus1940}, which is positive
definite and can be treated as quasi-probability density. There are three main
localization measures: $A$, the information entropy measure, $C$ the correlation
localization measure, and $nIPR$ the normalized inverse participation ratio.
As recently shown \cite{BatRob2013A,BLR2019}, they are all proportional to each
other (linearly related) and thus equivalent.  The energy spectra of the localized
chaotic eigenstates can be well described by the fractional power law level
repulsion,  $P(S)\propto S^{\beta}$, for small $S$, and $\beta \in [0,1]$: $\beta=0$
corresponds to the maximal localization and Poissonian statistics, while $\beta=1$
corresponds to the maximal extendedness (delocalization) and the RMT statistics.
It has been found that $\beta$ is a function of $A$, they are linearly related
in QKR and in the stadium billiard, as well as in the present work.
It is also an almost rational function of $\alpha$.

The local behaviour of $P(S)$ at small $S$
can be globalized by approximating it by the well known
Brody distribution \cite{Bro1973,Bro1981}, described by the following formula

\be \label{BrodyP}
P_B(S) = c S^{\beta} \exp \left( - d S^{\beta +1} \right), \;\;\; 
\ee
where by normalization of the total probability and the first moment we have

\be \label{Brodyab}
c = (\beta +1 ) d, \;\;\; d  = \left( \Gamma \left( \frac{\beta +2}{\beta +1}
 \right) \right)^{\beta +1}
\ee
with  $\Gamma (x)$ being the Gamma function. It interpolates the
exponential and Wigner distribution as $\beta$ goes from $0$ to $1$.
The corresponding gap probability is

\be \label{BrodyE}
E_B(S)  =  \frac{1}{\gamma (\beta +1)  } 
  Q \left( \frac{1}{\beta +1}, \left( \gamma S \right)^{\beta +1} \right).
\ee
where $\gamma=\Gamma \left(\frac{\beta +2}{\beta +1}\right)$ 
and $Q(a, x)$ is the incomplete Gamma function

\be \label{IGamma}
Q(a, x) = \int_x^{\infty} t^{a-1} e^{-t} dt.
\ee
Here the only parameter is $\beta$, the level repulsion exponent in
(\ref{BrodyP}), which measures the degree of localization of
the chaotic eigenstates: if the localization is maximally strong,
the eigenstates practically do not overlap in the phase space
(of the Wigner functions) and we find $\beta=0$ and Poissonian
distribution, while in the case of maximal extendedness (no localization)
we have $\beta=1$, and the RMT statistics of levels applies.
Thus, by replacing $E_{RMT}(S)$ with $E_B(S)$ we get the so-called
Berry-Robnik-Brody (BRB) distribution,
which generalizes the Berry-Robnik (BR) distribution such that the localization
effects are included \cite{BatRob2010}.
In this way the problem of describing the energy level statistics
is empirically solved. However, the theoretical derivation of
the Brody distribution for the localized chaotic states remains
an important open problem.

One  important theoretical plausibility argument by Izrailev in support of
Brody (or Brody-like)  intermediate level spacing distribution is that
the joint level distribution of Dyson circular ensembles can be extended
to noninteger values of the exponent $\beta$ \cite{Izr1990}.
The Izrailev distribution is a bit more complicated but has the feature of being
a better approximation for the GOE distribution at $\beta=1$.
However, recent numerical results show that Brody distribution 
is slightly better in describing real data
\cite{BatRob2010,BatRob2013A,ManRob2013,BatManRob2013}, 
and is simpler, which is the reason why we prefer and use it.

This paper is a continuation of our recent works on the
mixed-type billiard \cite{Rob1983,Rob1984,LozRob2018B},
classical and quantal. The role of the divided phase space and
of the localization effects of chaotic eigenstates
has been extensively studied in Refs. \cite{BatRob2010,BatRob2013A,BatRob2013B}. 

In the very recent papers on the classical dynamics in the
stadium billiard \cite{LozRob2018A} we have carefully investigated
the classical diffusion and transport properties, while
in Refs. \cite{BLR2018,BLR2019} we have performed a complete
analysis of the following quantal aspects: we have shown that $nIPR$ and
$A$ are equivalent, $A$ has a distribution on a compact
interval $[0,A_0]$, very well described by the beta distribution.
We have shown some representative Poincar\'e-Husimi functions
of various degrees of localization $A$. The mean value
$\mA$ is approximately a rational function of $\al$, the standard
deviation $\sigma$ of $A$ is analyzed, and we have shown that
the level repulsion exponent $\bet$ is a linear function of $\mA$,
and an almost rational function of $\al$, consistently with
the other properties. Finally, $\sigma$ seems to be a unique
function of $\bet$.

The purpose of the present paper is to carry out the same
complete analysis of the chaotic eigenstates in the mixed-type
billiard introduced in \cite{Rob1983,Rob1984}, showing that
all statistical properties of localized chaotic eigenstates
are universal, if the system's chaotic component is without pronounced
stickiness regions: The distribution $P(A)$ is beta distribution.
On the other hand, if the stickiness regions exist and are
pronounced, $P(A)$ is nonuniversal, it can have several maxima
(usually two) and each secondary maximum can be attributed to a
stickiness region.

The paper is organized as follows. In section \ref{sec2} we define
the billiard system, the Poincar\'e-Husimi functions,
introduce a method to separate regular and chaotic eigenstates,
and define the localization measures $A$ and $nIPR$,
and show that they are equivalent. In section \ref{sec3} we show the
dependence of the moments of $A$ on $\al$ and present some
typical Poincar\'e-Husimi functions. In section \ref{sec4}
we calculate the localization measures $A$ and their distribution
functions in various classical dynamical regimes. In section
\ref{sec5} we analyze the energy spectra and their statistical
properties (the level spacing distributions) as functions of
$\al$ in various classical dynamical regimes. In section \ref{sec6}
we draw the conclusions and discuss them in the context of further open
problems.

\section{The billiard system, definition of the Poincar\'e-Husimi
  functions, separation of regular and chaotic eigenstates,
  and the localization measures A and nIPR}
\label{sec2}

\subsection{The billiard system}
\label{subsec2.1}

The mixed-type billiard system  ${\cal B}$ in this paper has been introduced in Refs.
\cite{Rob1983,Rob1984}, and has been further studied by many others as a
model system, most extensively recently by Lozej and Robnik \cite{LozRob2018B}.
Its shape is defined by the complex  quadratic conformal map from
the unit circle $|z|=1$ in the $z$-plane onto the physical $w$-plane,

\be \label{lamdabilliard}
w = z + \lambda\; z^2,
\ee
where the family parameter $\la$ goes from $0$ to $1/2$: at $\la=0$ we
have the integrable circular billiard, for $0 < \la < 1/4$ it is a
convex shape having mixed-type phase space, in particular we have
the Lazutkin's caustics in the $w$-plane, and corresponding invariant curves in the
phase space. For $\la=1/4$ it is still convex but has a zero curvature
point at $z=-1$ and therefore (Mather's theorem) all
Lazutkin's tori are distroyed, allowing for
ergodicity, which, however does not yet occur, as we numerically
still find islands of stability. For $1/4 < \la <1/2$
it is nonconvex but still has a smooth boundary and very tiny islands
of stability (regular islands), while for $\la=1/2$ it has a cusp singularity
at $z=-1$ and has been proven by Markarian \cite{Markarian1993} to
be ergodic. Thus the system is an interesting and quite well
explored one-parameter family of billiards
going from the integrable circle billiard to a rigorously ergodic
and fully chaotic billiard, having mixed-type dynamics for the
intermediate values of $\la$. A recent very extensive survey of the
chaotic phase space has been performed by Lozej \cite{Lozej2019}.

For a 2D billiard the most natural coordinates in the phase space
$(s,p)$ are the arclength $s$ round the billiard boundary in
the mathematically positive sense (counterclockwise). $s\in [0,{\cal L}]$,
where ${\cal L}$ is the circumference, in our case starting at $s=0$ at the
point  $z=1$. The sine of
the reflection angle $\theta$ , which is the component of the unit velocity
vector tangent to the boundary at the collision point, equal to $p=\sin \theta$,
is the canonically conjugate momentum to $s$. These are the
Poincar\'e-Birkhoff coordinates. The bounce map $(s_1,p_1)
\rightarrow (s_2,p_2)$ is area preserving \cite{Berry1981}, and the phase portrait
does not depend on the speed (or energy) of the particle.

Quantum mechanically
we have to solve the stationary Schr\"odinger equation, which in a
billiard is just the Helmholtz equation 

\be \label{Helmholtz}
\Delta \psi + k^2 \psi =0
\ee
with the Dirichlet boundary conditions  $\psi|_{\partial {\cal B}}=0$.
The energy is $E=k^2$. The important quantity is
the boundary function 

\be  \label{BF}
u(s) = {\bf n}\cdot \nabla_{{\bf r}} \psi \left({\bf r}(s)\right),
\ee
which is the normal derivative of the wavefunction $\psi$ at the 
point $s$ (${\bf n}$ is the unit outward normal vector). 
It satisfies the integral equation

\be \label{IEBF}
u(s) = -2 \oint dt\; u(t)\; {\bf n}\cdot\nabla_{{\bf r}} G({\bf r},{\bf r}(t)),
\ee
where $G({\bf r},{\bf r'}) = -\frac{i}{4} H_0^{(1)}(k|{\bf r}-{\bf r'}|)$ is
the Green function in terms of the Hankel function $H_0^{(1)}(x)$. It is important
to realize that the boundary function $u(s)$ contains complete information
about the wavefunction at any point ${\bf r}$ inside the billiard by the equation

\be \label{utopsi}
\psi_m({\bf r})  = - \oint dt\; u_m(t)\; G\left({\bf r},{\bf r}(t)\right).
\ee
Here $m$ is just the index (sequential quantum number) of the $m$-th eigenstate.

\subsection{The Poincar\'e-Husimi functions}
\label{subsec2.2}

Now we go over to the quantum phase space. We can calculate the Wigner
functions \cite{Wig1932} based on $\psi_m({\bf r})$. However, in billiards it is advantageous to
calculate the Poincar\'e-Husimi functions. The Husimi functions \cite{Hus1940} are
generally just Gaussian smoothed Wigner functions. Such smoothing makes
them positive definite, so that we can treat them somehow as quasi-probability 
densities in the quantum phase space, and at the same time we eliminate the
small oscillations of the Wigner functions around the zero level, which do
not carry any significant physical contents, but just obscure the picture.
Thus, following  Tualle and Voros \cite{TV1995} and B\"acker et al
\cite{Baecker2004}, we introduce \cite{BatRob2013A,BatRob2013B} 
the properly ${\cal L}$-periodized coherent states
centered at $(q,p)$, as follows

\bea \label{coherent}
c_{(q,p),k} (s) & =  & \sum_{m\in {\bf Z}} 
\exp \{ i\,k\,p\,(s-q+m{\cal L})\}  \times \\ \nonumber
 & \exp & \left(-\frac{k}{2}(s-q+m{\cal L})^2\right). 
\eea
The Poincar\'e-Husimi function is then defined as the absolute square
of the projection of the boundary function $u(s)$ onto the coherent
state, namely

\be \label{Husfun}
H_m(q,p) = \left| \int_{\partial {\cal B}} c_{(q,p),k_m} (s)\;
u_m(s)\; ds \right|^2.
\ee

\subsection{The localization measures   A and nIPR}
\label{subsec2.3}

The {\em entropy localization measure} of a {\em single
eigenstate}  $H_m(q,p)$, denoted by $A_m$ is defined as

\be \label{locA}
A_m = \frac{\exp I_m}{N_c},
\ee
where

\be  \label{entropy}
I_m = - \int dq\, dp \,H_m(q,p) \ln \left((2\pi\hbar)^f H_m(q,p)\right)
\ee
is the information entropy.  Here $f$ is the number of degrees
of freedom (for 2D billiards $f=2$, and for surface of section it is
$f=1$) and $N_c$ is a number of cells on the 
classical chaotic domain, $N_c=\Omega_c/(2\pi\hbar)^f$, where
$\Omega_c$ is the classical phase space volume of the classical chaotic component.
In the case of the
uniform distribution (extended eigenstates) $H=1/\Omega_C={\rm const.}$
the localization measure is $A=1$, while in the case of the strongest localization
$I=0$, and $A=1/N_C \approx 0$.
The Poincar\'e-Husimi function $H(q,p)$
(\ref{Husfun}) (normalized) was calculated on the grid points $(i,j)$
in the phase space $(s,p)$,  and
we express the localization measure in terms of the discretized function.
In our numerical calculations we have put $2\pi\hbar=1$, and
thus we have $H_{ij}=1/N$, where $N$ is the number of grid points,
in case of complete extendedness, while for maximal localization
we have $H_{ij}=1$ at just one point, and zero elsewhere.
In all calculations
have used the grid of $400\times 400$ points, thus $N = 160000$.

\begin{figure}[H]
  \centering
  \includegraphics[width=9cm]{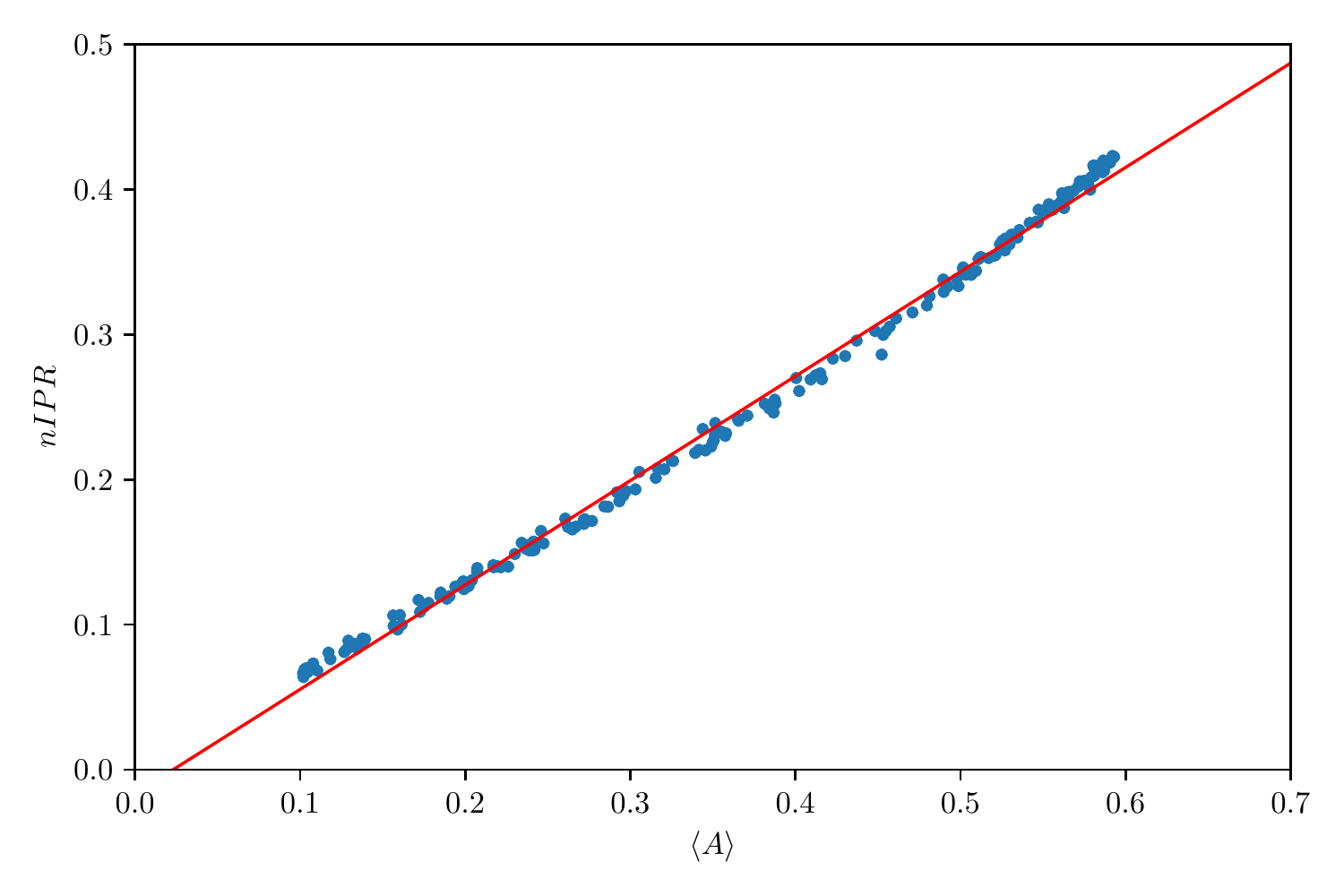}
  \caption{The normalized inverse participation ratio as a localization measure,
    as a function of $A$. They are linearly related and thus equivalent. We have used the
  cahotic states with $M\ge M_t=0.5$. The slope is $0.72$ and the intercept $-0.017$.} 
  \label{nIPRofA}
\end{figure}

As mentioned in the introduction, the definition of localization measures
can be diverse, and the question arises to what extent are the results
objective and possibly independent of the definition. Indeed, in reference
\cite{BatRob2013A}, it has been shown that $A$ and $C$ (based on the corelations)
are linearly related and thus equivalent. Moreover, we have introduced also
the normalized inverse participation ratio $nIPR$, defined as follows

\be \label{nIPR}
nIPR = \frac{1}{N} \frac{1}{\sum_{i,j} H_{ij}^2},
\ee
for each individual eigenstate $m$. However, because we expect fluctutaions
of the localization measures even in the quantum ergodic regime (due to the scars etc),
we must perform some averaging over an ensemble of eigenstates, and for this we have
chosen $100$ consecutive eigenstates. Then, by doing this for all possible data for the
the billiard at various $\la$ and $k$, we ended up with the result that
the $nIPR$ and $A$ are linearly related and thus also equivalent, as shown in
Fig. \ref{nIPRofA}. This is in perfect agreement with the most recent results for the
stadium billiard \cite{BLR2019}, and thus we believe that it is generally true,
independent of a specific model system.

In the following we shall use exclusively $A$ as the measure
of localization.

\subsection{Introducing the distribution of the localization measure A}
\label{subsec2.4}

The central object of interest in this paper is the distribution $P(A)$
of the localization measures $A_m$ of the chaotic eigenstates
within a certain interval of 2000 consecutive
even-parity eigenstates indexed by $m$, around some central value $k_0$.
We have done this for 18 different values of $\la$ and for each $\la$
for 9 to 12 different values of $k_0$.
Each distribution function $P(A)$, generated by the segment of chaotic
eigenstets within the stretch of 
2000 consecutive values $A_m$, is defined on a compact interval $[0,A_0]$.
Ideally, according to the above Eqs. (\ref{locA},\ref{entropy}), the maximum
value of $A$ should be $1$, if the Husimi function were entirely and uniformly
extended. However, this is never the case, as the Husimi functions have zeros
and oscillations, and thus we must expect a smaller maximal value, smaller  than $1$,
which in addition might vary from case to case, depending on $k$ and the grid size.
As long as we do not have a theoretical prediction for $A_0$, we must proceed
empirically. Therefore we have checked several values of $A_0$ around $A_0=0.7$,
and found that the latter value is the best according to several criteria.
See also the discussion at the end of section \ref{sec4}.

We shall look at the moments of $P(A)$, namely

\be \label{locA2}
\left< A \right>= \int_0^{A_0}A\, P(A) \,dA, \,\,\,
\left< A^2 \right>=  \int_0^{A_0}A^2\, P(A) \,dA,
\ee
and the standard deviation 

\be \label{locA3}
\sigma =\sqrt{\left< A^2 \right> - \left< A \right>^2}.
\ee
For the numerical calculations of the eigenfunctions $\psi_m({\bf r})$
and the corresponding energy levels $E_m=k_m^2$  we have used the
Vergini-Saraceno method  \cite{VerSar1995}. Also, we have calculated only the
even symmetry class of solutions. 

\subsection{The separation of regular and chaotic eigenstates}
\label{subsec2.5}

Now the classification
of eigenstates can be performed by their projection onto the classical
surface of section. As we are very deep in the semiclassical regime
we do expect with probability one that either an eigenstate is
regular or chaotic, with exceptions having measure zero, ideally.
To automate this task we have ascribed to each point on the grid
a number $K_{i,j}$ whose value is either $+1$ if the grid point lies
within the classical chaotic region or $-1$ if it belongs to a 
classical regular region. Technically, this has been done as follows.
We have taken an initial condition in the chaotic region, and iterated 
it up to about $10^{10}$ collisions, enough for the convergence
(within certain very small distance).
Each visited cell $(i,j)$ on the grid has then been assigned value 
$K_{i,j} = +1$, the remaining ones were assigned the value -1.

The Poincar\'e-Husimi function $H(q,p)$
(\ref{Husfun}) (normalized) was calculated on the grid points and the overlap
index $M$ was calculated according to the definition

\be \label{indexM}
M = \sum_{i,j} H_{i,j}\; K_{i,j}.
\ee
In practice, $M$ is not exactly $+1$ or $-1$, but can have a value
in between. The reasons are two, first the finite discretization
of the phase space (the finite size grid), and second,  the
finite wavelength (not sufficiently small effective Planck constant,
for which we can take just $1/k_j$). If so, the question is, where
to cut the distribution of the $M$-values, at the threshold value $M_t$,
such that all states with $M<M_t$ are declared regular and 
those with $M>M_t$ chaotic.

There are two natural criteria: {\bf (I)} {\em The classical criterion:} 
the threshold value $M_t$ is chosen such that we have exactly
$\rho_1$ fraction of regular levels and $\rho_2=1-\rho_1$ of
chaotic levels. {\bf (II)} {\em The quantum criterion:} we choose $M_t$
such that we get the best possible agreement of the chaotic
level spacing distribution with the Brody distribution (\ref{BrodyP}),
which is expected to capture the dynamical localization effects
of the chaotic eigenstates.  However, when we wanted to make sure
that only chaotic eigenstates are being used, we have chosen $M_t=0.5$.

\section{ Moments of A  and examples of Poincar\'e-Husimi functions}
\label{sec3}

The system parameter governing the localization phenomenon
$\al = t_H/t_T$, as introduced in Eq. (\ref{alpha}),
in a quantum billiard described by the Schr\"odinger equation
(Helmholtz equation) Eq. (\ref{Helmholtz}), becomes

\begin{equation} \label{A7}
\alpha = \frac{2k}{N_T}.
\end{equation}
where $N_T$ is the discrete classical transport time, that is the
characteristic number of collisions of the billiard particle
necessary  for the global spreading of the ensemble of uniform in $s$
initial points at zero momentum in the momentum space.
This quantity $N_T$ can be defined in various ways as discussed
in references \cite{LozRob2018B,BLR2018,BatRob2013A,BatRob2013B}, where
the derivation of $t_T$, $N_T$ and $\al$ is given. Unlike the stadium billiard,
where the diffusion can be very slow and well described by the
exponential approach to the equilibrium value (uniformly spread
ensemble in the entire phase space), in the present billiard the
classical spreading (transport) is not described by a diffusion law,
but still can be well described by the criterion
of the second moment $\left< p^2 \right>$ reaching a certain
fraction (percentage) of the asymptotic, maximal, value. Indeed, in spite
of some arbitrariness of this definition, we found it sound,
as the final results do not qualitatively depend on the choice
of the criterion, only some parameters change their values, as we
shall see below. In Table I we give the values of $N_T$ according to
the four different criteria (50\%, 70\%, 80\% and 90\%), for a variety
of values of $\la$, which we consider in this paper in the sections
to follow.

\begin{table}
  \center
  \begin{tabular}{ | p{1.2cm} | p{1.2cm}  | p{1.2cm} | p{1.2cm} | p{1.2cm}|}
  \hline
  \multicolumn{5}{|c|}{The classical transport time $N_T$}\\
  \hline
  $\lambda$ &          $90\%$  &        $80\%$  &   $70\%$  & $50\%$\\
  \hline
   0.135 &   48218 &  13893 &    6444 &   2325 \\ \hline
   0.140 &   26830 &   7992 &    3750 &   1227 \\ \hline
   0.145 &   21284 &   5089 &    2501 &    936 \\ \hline
   0.150 &   11431 &   3289 &    1579 &    534 \\ \hline
   0.155 &    6134 &   2164 &    1103 &    405 \\ \hline
   0.160 &    3981 &   1332 &     706 &    264 \\ \hline
   0.165 &    2506 &    908 &     509 &    205 \\ \hline
   0.170 &    1763 &    678 &     440 &    182 \\ \hline
   0.175 &    1569 &    592 &     327 &    127 \\ \hline
   0.180 &    1257 &    480 &     248 &     86 \\ \hline
   0.185 &     737 &    319 &     177 &     62 \\ \hline
   0.190 &     542 &    258 &     152 &     55 \\ \hline
   0.200 &     314 &    170 &     106 &     47 \\ \hline
   0.210 &     287 &    138 &      85 &     37 \\ \hline
   0.220 &     192 &     92 &      55 &     23 \\ \hline
   0.230 &     147 &     73 &      44 &     17 \\ \hline
   0.240 &     106 &     52 &      29 &      9 \\ \hline
   0.250 &      77 &     40 &      21 &      7 \\ \hline
\end{tabular}
\caption{The classical transport time $N_T$ in units of the number of collisions for the
  criteria of certain percentage of the asymptotica maximal value of the second
  moment of momentum distribution, starting from zero (Dirac delta dsitribution), calculated
numerically for various values of $\la$.}
\end{table}

The condition for the occurrence of dynamical localization
$\alpha \le 1$ is now expressed in the inequality

\begin{equation} \label{A8}
k \le \frac{N_T}{2},
\end{equation}
although the empirically observed transitions are not at all sharp with $\al$.

\begin{figure}[H]
  \centering
  \includegraphics[width=9cm]{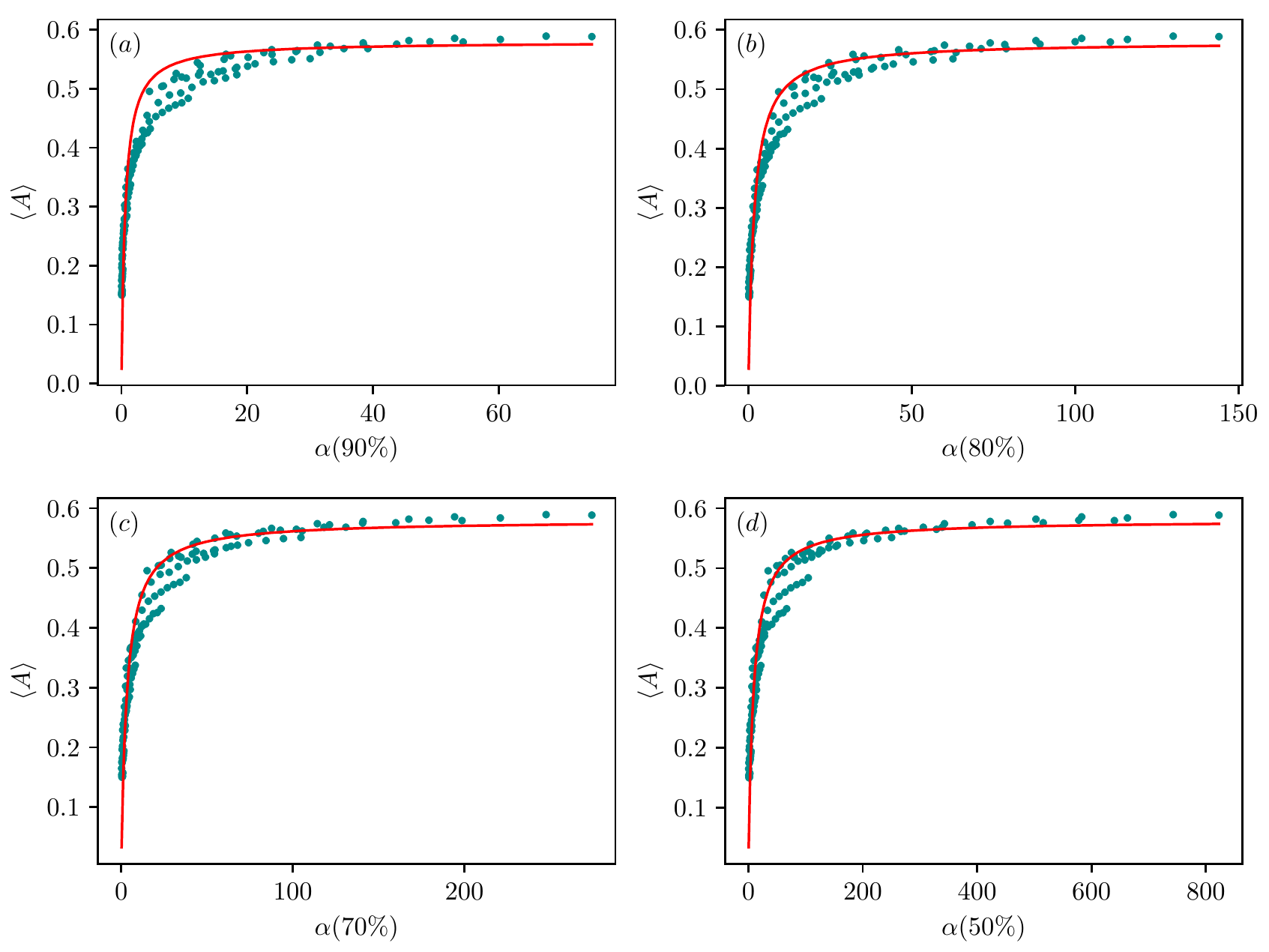}
  \caption{The mean entropy localization measure $\mA$ for variety of
    $\la$ and energies $E=k^2$, as a function of $\al$ fitted
  by the function (\ref{Avsalphaeq}), based on $N_T$ from the 
  Table I, with $A_{\infty}=0.58$ and $s=1.70,\;0.57,\;0.30,\; 0.11$, for
  (a-d), respectively.
  Qualitatively it is very similar to the stadium \cite{BLR2018,BLR2019}.} 
  \label{lambdaAvsalpha}
\end{figure}

In Fig. \ref{lambdaAvsalpha}  we show the dependence of $\mA$ on $\al$,
where $\al$ is calculated using $N_T$ from Table I. Moreover, in order to compare
and well define the $A$ for various values of $\la$, we have to divide these
values by the relative size (area) $\chi_C$ of the largest chaotic component.
Otherwise, we would see different values of $A$ even for entirely  extended states,
due to the different size of the classical chaotic components.
Therefore, in the case of full extendeness they obtain all
the same value $A_0$, for which it turns out empirically that $A_0=0.7$ is the best choice.
The table of the $\chi_C$ values was published in the paper \cite{LozRob2018B}.
Like in the stadium \cite{BLR2018,BLR2019} the transition
from strong localization of small $\mA$ and $\al$ to the complete delocalization
$\al \gg 1$ is quite smooth, over almost two decadic orders of magnitude.
As we see, $\mA$ is approximately fitted by a rational function of $\al$,  namely

\be  \label{Avsalphaeq}
\mA = A_{\infty} \frac{s\al}{1 +s \al},
\ee
where the values of the parameters are $A_{\infty}=0.58$ and $s=1.70,\;0.57,\;0.30,\; 0.11$, for
(a-d), respectively.

\begin{figure}[H]
  \centering
  \includegraphics[width=9cm]{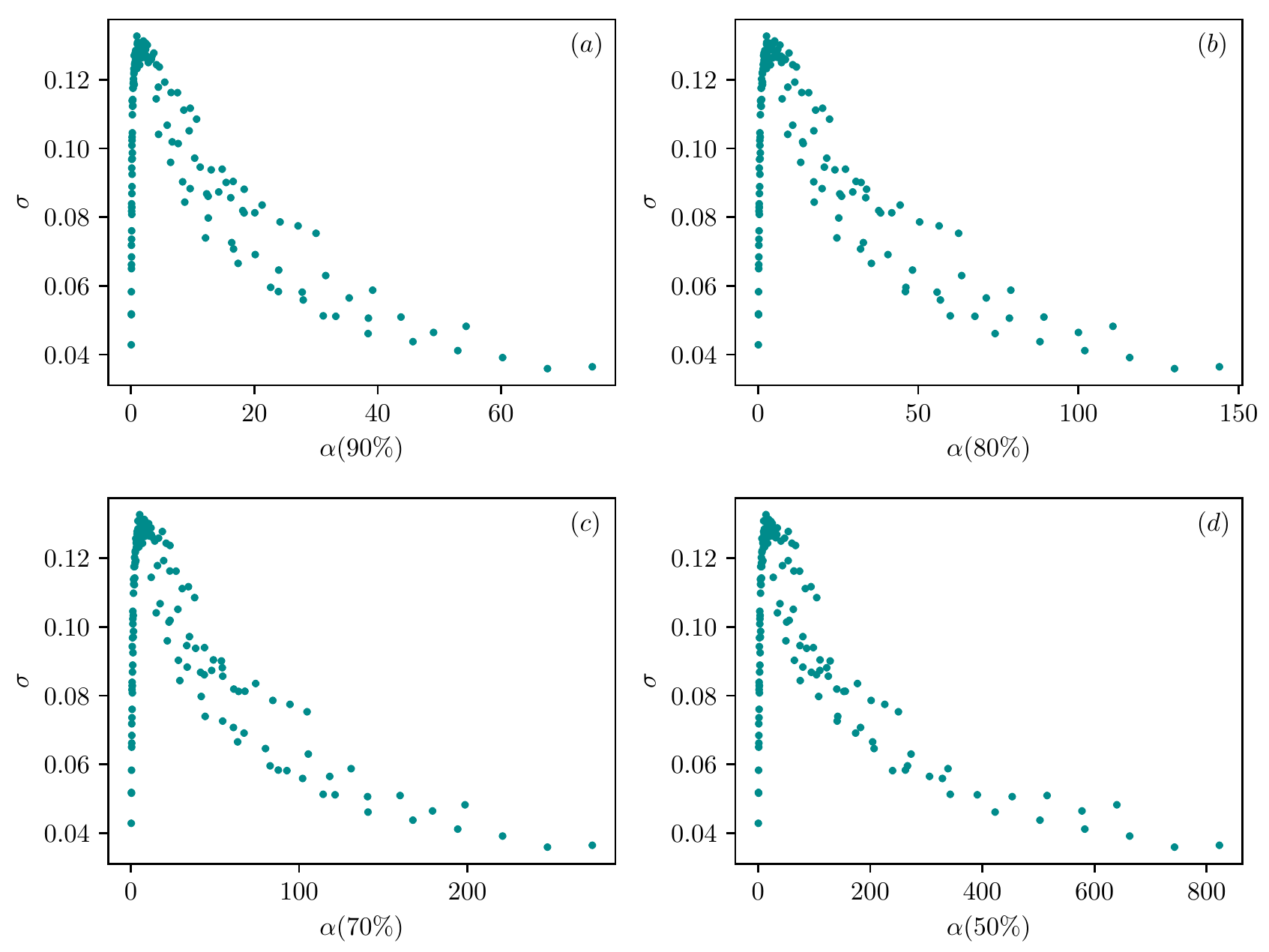}
  \caption{The standard deviation $\sigma$ as a function of the
    $\al$ for variety of different shapes $\la$ and energies $E=k^2$.
  Qualitatively it is very similar to the stadium \cite{BLR2018,BLR2019}.}
  \label{lambdasigmaofalpha}
\end{figure}

In Fig. \ref{lambdasigmaofalpha} we show the dependence of $\sigma$ defined in
Eq. (\ref{locA3}) upon $\al$ also using $N_T$ from Table I.

We see that while $\mA$ is
a monotonically increasing function of $\al$ (Fig. \ref{lambdaAvsalpha}),
the standard deviation $\sigma$ starts at zero,
is small for small $\al$, but rises sharply,
and reaches some maximum at about $\al\approx 10$, and then decreases
very slowly at large values of $\al$. Thus both, the very strongly
localized eigenstates, mimicking invariant tori, and the
entirely delocalized (ergodic) eigenstates have 
small spreading $\sigma$ around the mean value $\mA$. According to the
quantum ergodic theorem of Shnirelman \cite{Shnirelman1974}
$\sigma$ should tend to zero when $\al\rightarrow \infty$,
and rescaled $\mA\rightarrow 1$, but the transition
to that regime might be very slow as suggested by Fig. \ref{lambdasigmaofalpha}.
In this limit $P(A)$ must become the Dirac delta function peaked at $A_0$.
However, it is very difficult to quantify this approach quantitatively,
as at large $\al$ we have very few physically reliable data
points, so it is too early to draw any definite conclusion
about the asymptotic behavior at $\al \rightarrow \infty$.
More numerical efforts are needed, currently not feasible.

\begin{figure*}
  \begin{centering}
    \includegraphics[width=1\textwidth]{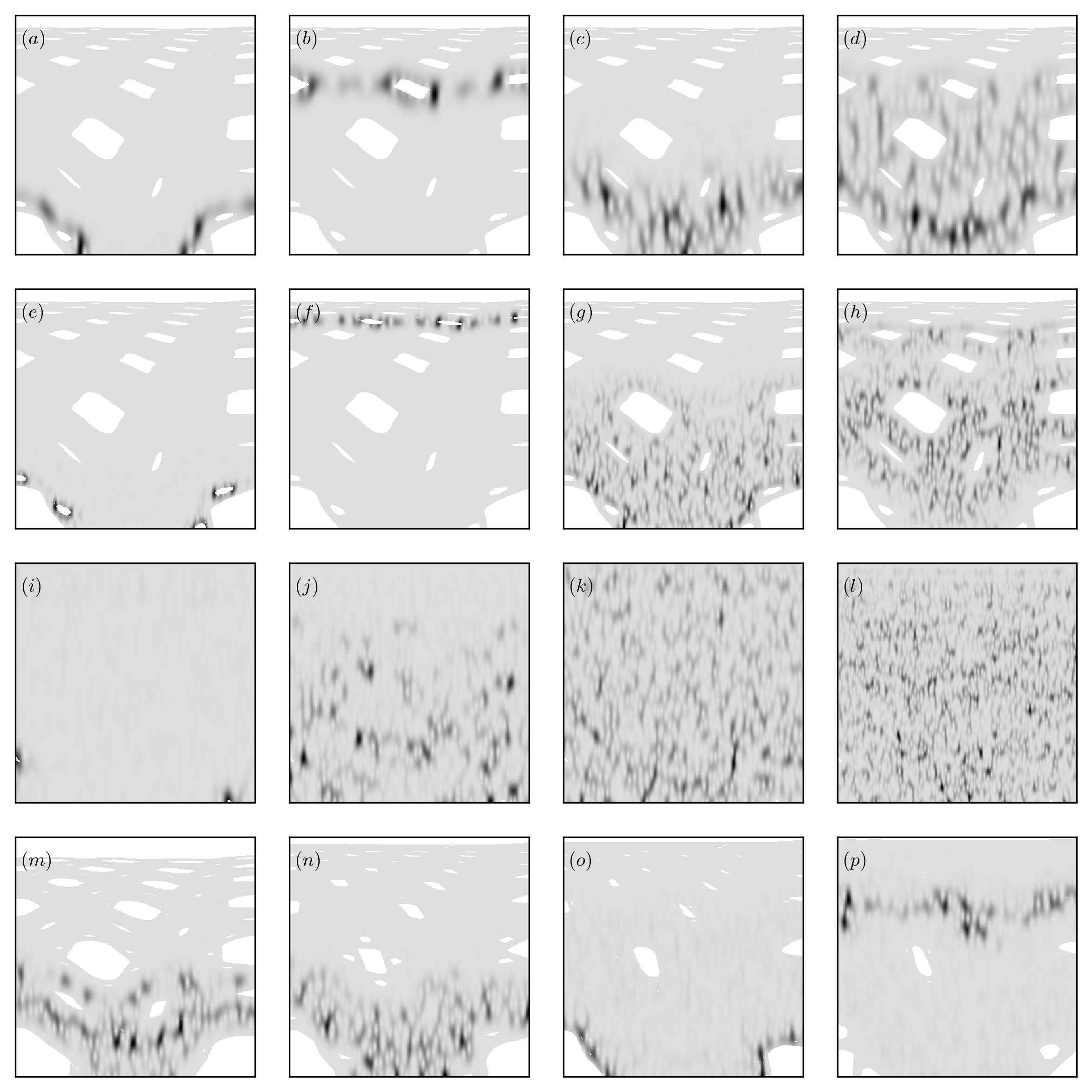}
    \par\end{centering}
  \caption{We show plots of Poincar\'e-Husimi functions for a
    representative selection of chaotic eigenstates:
    Row (a-d): $\la=0.15$, $k=$ $642.5819901$,  $634.10919221$,  $635.84172768$,  $634.60937942$;
    Row (e-h): $\la=0.15$, $k=$ $2598.54009629$, $2598.5812614$, $2598.7038606$, $2601.4614965$;
    Row (i-l): $\la=0.25$, $k=$ $640.68503066$, $1197.17988119$, $1477.59419041$, $3719.0489485 $;
    Row (m-p): $\la=0.14$, $k=1196.82174307$; $\la=0.16, k= 1196.83132015$;
    $\la=0.17$, $k=1196.87057788$; $\la=0.18$, $k=1197.13132098$.
    }
  \label{Husimipic}
\end{figure*}

The Poincar\'e-Husimi functions describe the structure of the localized
chaotic eigenstates. In Fig. \ref{Husimipic} we show some selection of
typical Poincar\'e-Husimi functions for various values of $\la$ and $k$,
and the corresponding $\al$. We show only the upper right quadrant 
$s \in [0,{\cal L}/2]$, $p\in [0,1]$  of the
classical phase space, as due to the symmetries (reflection symmetry
and the time reversal symmetry) all four quadrants are equivalent.

At large $\la=0.25$ (almost ergodic case) and fixed $k$, we have small $N_T$ and according
to Eq.(\ref{A7}) $\al \gg 1$, we observe mainly ergodic eigenstates,
in agreement with the quantum ergodic theorem \cite{Shnirelman1974}, that is
fully extended states, exemplified in (k,l). Nevertheless, there are some exceptions,
asymptotically of measure zero, where we observe partial localization,
as shown in (j). Also, some Poincar\'e-Husimi functions can be associated
with small stability islands around a stable classical orbit exempliifed by (i).
Moreover, there can be strongly localized states corresponding
to the scaring around and along an unstable periodic orbit in the chaotic sea.
More precisely, the area of scars of eigenfunctions
$\psi_m({\bf r})$ goes to zero, and the relative number of scarred states
goes to zero as $\hbar\rightarrow 0$ or $m\rightarrow \infty$ \cite{Heller1984}.

As we decrease $\la=0.15$ and $\al$, thereby increasing $N_T$, the degree of localization
increases, thus $\mA$ is decreasing as shown in (a-h). At other values of
$\la=0.14,\; 0.16,\;  0.17,\; 0.18$ of strongly pronounced mixed-type phase space
we see localized states exemplified in (m-p).

\section{The distributions of the localization measures  A}
\label{sec4}

\begin{figure*}
  \begin{centering}
    \includegraphics[width=1\textwidth]{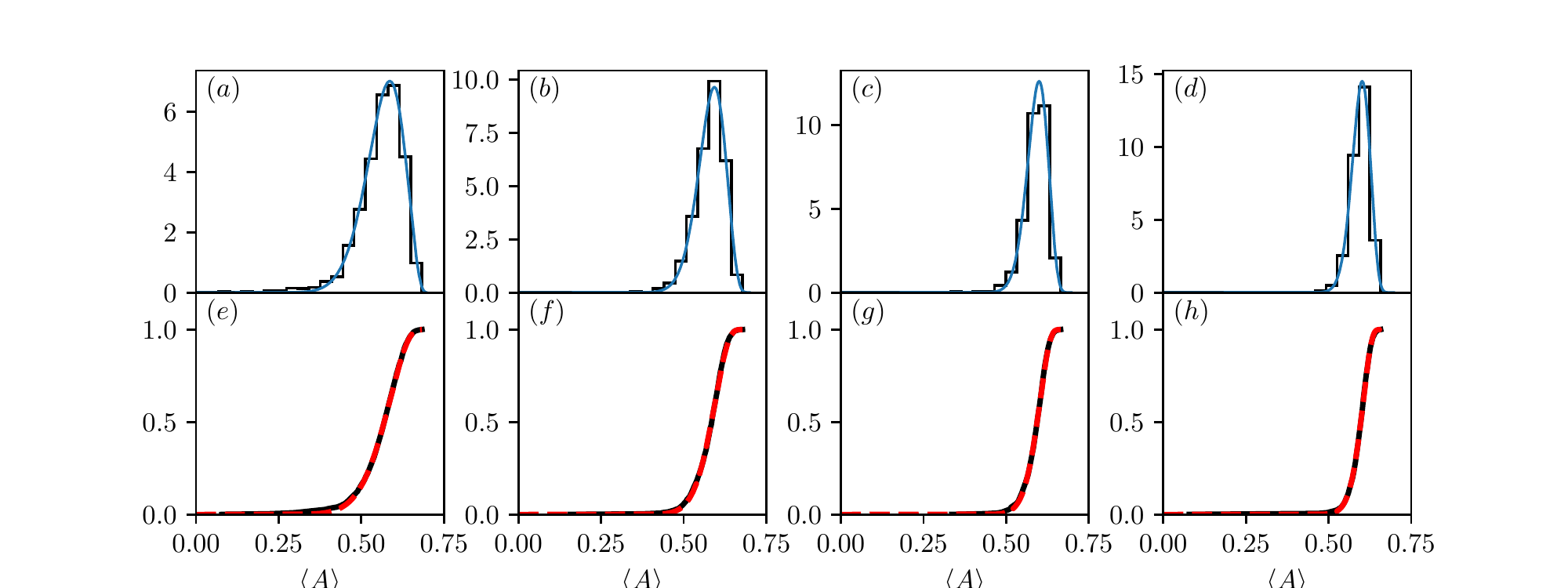}
    \par\end{centering}
  \caption{The distributions $P(A)$ of the entropy localization measure $A$
    for $\la=0.25$ and various $k_0$ (from (a) to (d)): $640,\; 1480,\; 2600,\; 3720$. [Colors online: black are data, blue is the best fit.]
    In (e-h) are the corresponding cumulative distributions (\ref{cumPA}).
[Colors online: black are data, red dashed is the best fit.]
    It is seen that
    the beta distribution fit is perfect. $A_0=0.7$  and the $(a,b)$ parameters of
    the best fit beta distribution are from (a) to (d): $(16.57,3.22),\; (30.68,5.55),\; (50.89,8.51),\; (66.61,10.87)$. }
  \label{histo1}
\end{figure*}

\begin{figure*}
  \begin{centering}
    \includegraphics[width=1\textwidth]{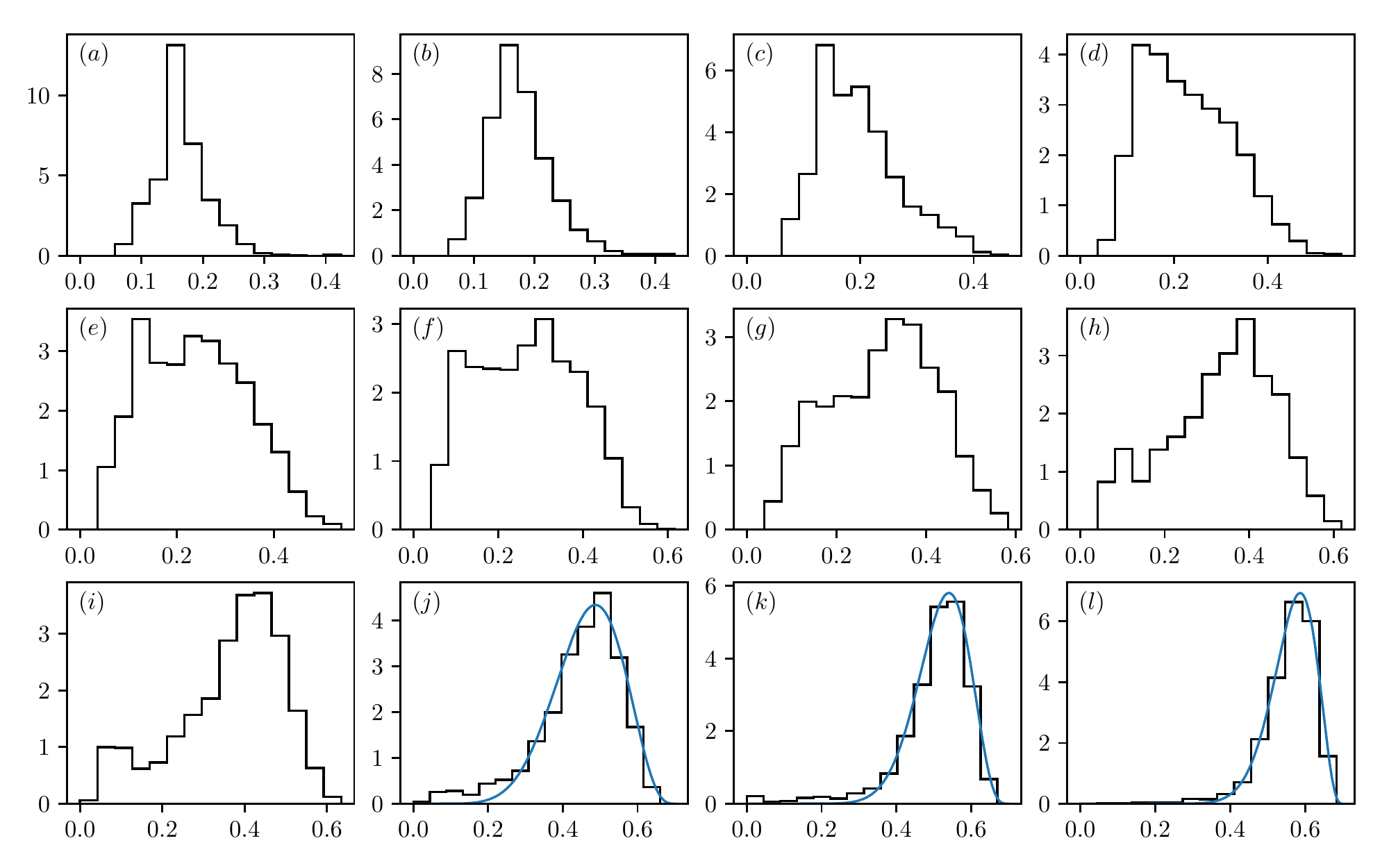}
    \par\end{centering}
  \caption{The distributions $P(A)$ of the entropy localization measure $A$
    for $k_0=640$ and various $\la$ (from (a) to (l)):
    $0.135,\;0140,\;0.145,\;0.150,\;0.155,\;0.160,\;0.165,\;0.170,\;0.180,\;0.200,\;
    0.220,\; 0.250$. The last three histograms (j-l) are well fitted by the beta
    distribution with $A_0=0.7$ and the $(a,b)$ parameter values: $(7.56,3.30),\; (13.16,3.89),\; (16.09,3.12)$.
    }
  \label{histo2}
\end{figure*}

\begin{figure*}
  \begin{centering}
    \includegraphics[width=1\textwidth]{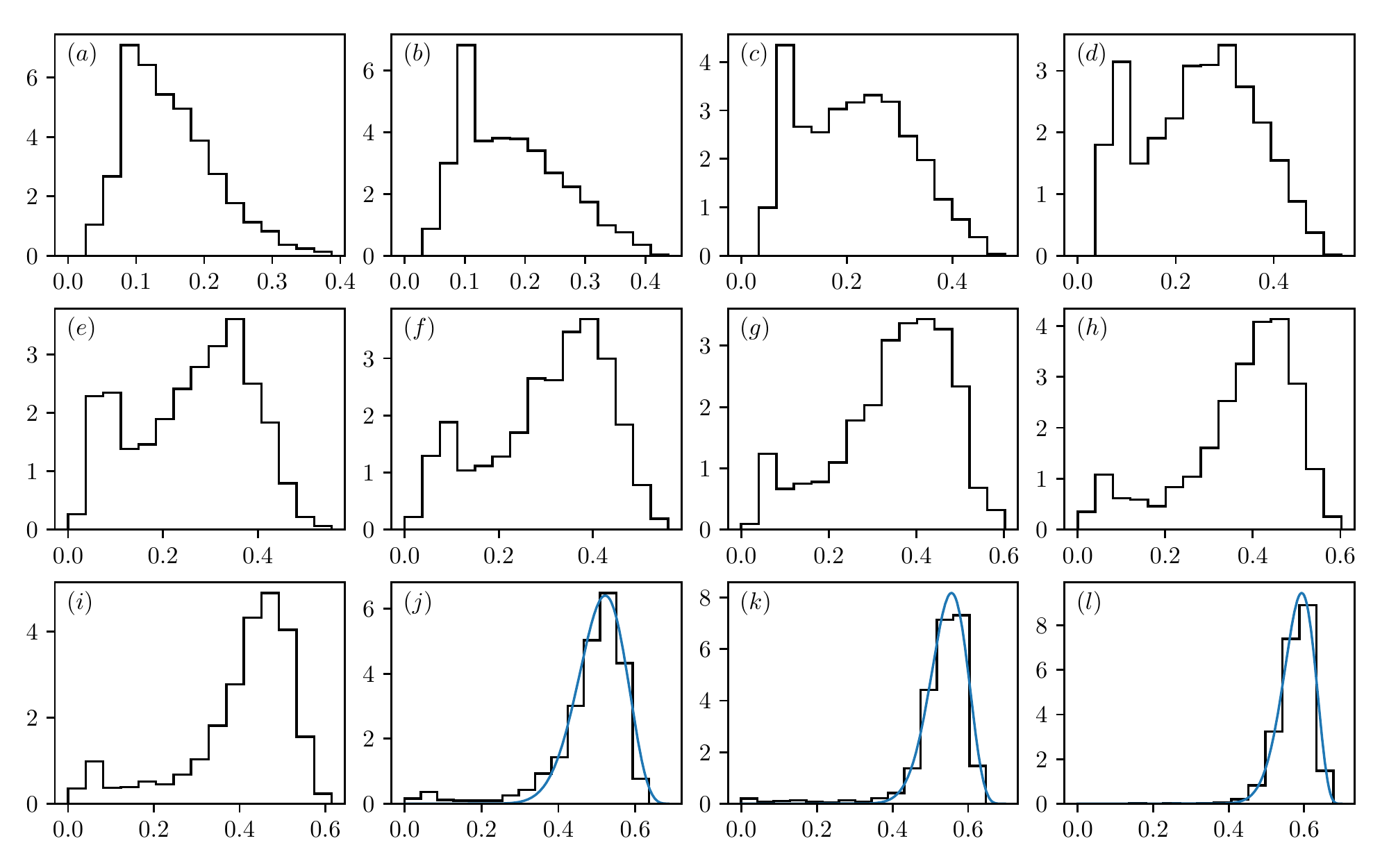}
    \par\end{centering}
  \caption{The distributions $P(A)$ of the entropy localization measure $A$
    for $k_0=1480$ and various $\la$ (from (a) to (l)):
    $0.135,\;0140,\;0.145,\;0.150,\;0.155,\;0.160,\;0.165,\;0.170,\;0.180,\;0.200,\;
    0.220,\; 0.250$. The last three histograms (j-l) are well fitted by the beta
    distribution with $A_0=0.7$ and the $(a,b)$ parameter values: $(16.93,5.79),\; (25.80,6.69),\; (29.36,5.27)$.
    }
  \label{histo3}
\end{figure*}

\begin{figure*}
  \begin{centering}
    \includegraphics[width=1\textwidth]{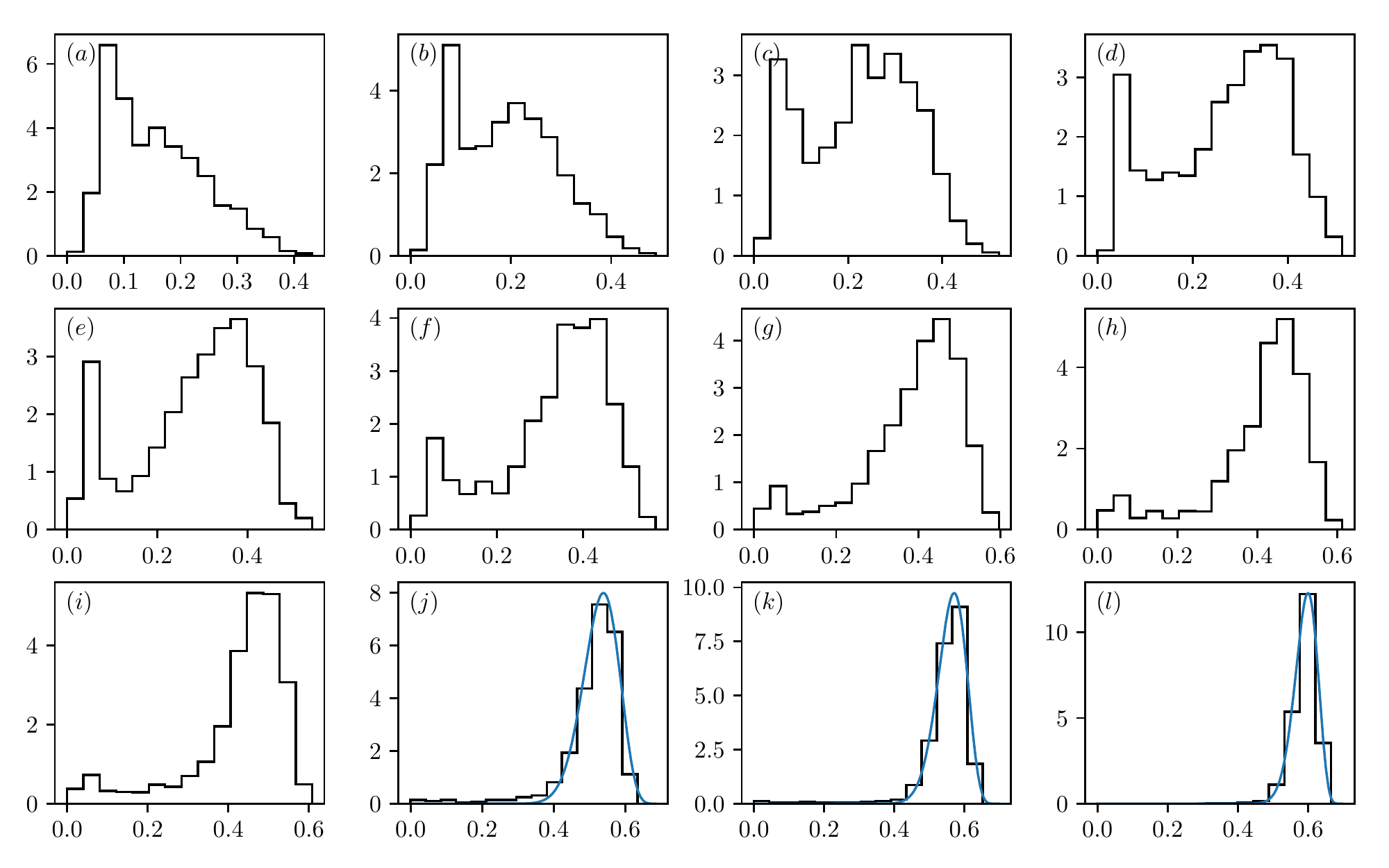}
    \par\end{centering}
  \caption{The distributions $P(A)$ of the entropy localization measure $A$
    for $k_0=1480$ and various $\la$ (from (a) to (l)):
    $0.135,\;0140,\;0.145,\;0.150,\;0.155,\;0.160,\;0.165,\;0.170,\;0.180,\;0.200,\;
    0.220,\; 0.250$. The last three histograms (j-l) are well fitted by the beta
    distribution with $A_0=0.7$ and the $(a,b)$ parameter values: $(25.86,7.74),\; (35.01,7.92),\; (48.46,8.14)$.
    }
  \label{histo4}
\end{figure*}

In this section we present the central results of this paper, namely the
distribution functions of the localization measures $A$. It is found,
according to our expectation, that in the almost ergodic case with large
$\la$ such as e.g. $\la=0.25$, we find  that each distribution can be  very
well characterized and described
by the so-called {\em beta distribution}

\be  \label{betadistr}
P(A) = C A^a (A_0-A)^b,
\ee
where $A_0$ is the upper limit of the interval $[0,A_0]$ on which $P(A)$ is defined,
and the two exponents $a$ and $b$ are positive  real numbers, while $C$ is the
normalization constant such that $\int_0^{A_0} P(A)\,dA = 1$, i.e.

\be \label{C}
C^{-1} = A_0^{a+b+1} B(a+1,b+1),
\ee
where $B(x,y) = \int_0^1 t^{x-1} (1-t)^{y-1} dt$ is the beta function.
We shall also use the cumulative distribution defined as

\be \label{cumPA}
W(A) = \int_0^A P(x)\, dx.
\ee
Thus we have

\be \label{mA}
\mA = A_0 \frac{a+1}{a+b+3},
\ee
and for the second moment

\be \label{2mA}
\left< A^2 \right>  = A_0^2 \frac{(a+2)(a+1)}{(a+b+4)(a+b+3)}
\ee
and therefore for the standard deviation  $\sigma$  (\ref{locA3}),

\be \label{sigmaA}
\sigma^2  = 
A_0^2 \frac{(a+2)(b+2)}{(a+b+4)(a+b+3)^2},
\ee
such that asymptotically  $\sigma \approx A_0 \frac{\sqrt{b+2}}{a}$ when
$a\rightarrow \infty$. Whenever we compare $A$ from different $\la$,
we have divided $A$ by the relative fraction of the chaotic component
in the classical phase space, denoted by $\chi_C$, as computed and
listed in the table of Ref. \cite{LozRob2018B}.
In the figures \ref{histo1}, \ref{histo2}, \ref{histo3}, \ref{histo4}
we show a selection of typical distributions $P(A)$.
In all cases for $A_0$ we have chosen the empirically best
value $A_0=0.7$. By $k_0$ we denote the mean
value of $k$ intervals on which we calculate the 2000 successive eigenstates,
from which we extract the chaotic ones, by choosing an appropriate value of
$M_t$, always $M_t=0.5$, in order to make sure that we collect chaotic states.
In addition, it should be noted that losing a few chaotic states, which
can happen, does not affect the result for $P(A)$ in any significant way.
It should be noted that the statistical significance is very high,
which has been carefully checked by using a (factor 2) smaller number of objects
in almost all histograms, as well as by changing the size of the boxes.

The limiting case $a \rightarrow \infty$ in Eqs.(\ref{mA},\ref{sigmaA}) comprising the
fully extended states in the limit $\al \rightarrow\infty$ shows that the
distribution tends to the Dirac delta function peaked at $A_0$,
thus $\sigma=0$ and $P(A)=\delta(A-A_0)$, in agreement with Shnirelman's theorem
\cite{Shnirelman1974}.

The Fig. \ref{histo1} clearly shows that the fit by the beta distribution (\ref{betadistr})
is excellent in case of $\la=0.25$,
typical for the ergodic regimes with $\al \gg 1$, as observed
also in the stadium in Ref. \cite{BLR2019}.
The qualitative trend from strong localization to weaker
localization or even complete extendedness (ergodicity) with increasing $k_0$
is clearly visible.

In the Fig. \ref{histo2} we show the distribution $P(A)$ for $k_0=640$ and several
values of the shape parameter $\la$. Here $A$ is normalized by dividing it with $\chi_C$.
The behaviour that we see is physically very interesting and statistically
significant, as we checked carefully, but is not universal, as it depends on the
structure of the chaotic component in the classical phase space and on
the size and intensity of the stickiness regions. Each stickiness region is
expected to support a local maximum in $P(A)$, although this phenomenon
is more strongly exhibited in the lemon billiards \cite{LLR2019}. Only at large $\la$, when
$\al \gg 1$, we see the approach to the beta distribution characteristic
of the ergodic regime without stickiness on the underlying chaotic region,
exemplified by (j-l), where $\sigma$ decreases towards the limit $\sigma=0$.
Similar observations apply to Figs. \ref{histo3} and \ref{histo4}.

We should stress that there is of course some arbitrarines in defining $A_0$,
so long as we do not have a theoretical prediction for its value. 
So far we have taken $A_0=0.7$, but nevertheless tried also the choice
of $A_0$ being the largest member $A$ in each histogram, and found no
significant qualitative changes, but only minor quantitative differences.
In both cases $a$ and $b$ are {\em not}
unique functions of $\al$, while $\langle A \rangle$ and $\sigma$ might be
unique functions of $\al$ as demonstrated in Figs. \ref{lambdaAvsalpha} and
\ref{lambdasigmaofalpha}, and approximated by a fit in Eq. (\ref{Avsalphaeq}).

\section{Implications of localization for the spectral statistics of chaotic eigenstates}
\label{sec5}

To get a good estimate of $\beta$ we need many more levels (eigenstates)
than in calculating $\mA$. 
The parameter $\beta$ was computed for 18 different values of the parameter
$\lambda$ as given in the Table I, and for
12 intervals in $k$ space: $(k_i,k_{i+1})$ where $k_i = 500 + 280\,i$
and $i\in[0,1,\dots, 11]$.
This is $18\times 12= 216$ values of $\beta$ altogether. More than $4\times10^6$
energy levels were computed for each $\lambda$. The size of the intervals
in $k$ was chosen to be maximal and such that the BRB (Berry-Robnik-Brody)
distribution gives a good fit to the level spacing distributions of the levels
in the intervals, meaning that $\bet$ is well defined.

For each $\beta(\lambda,(k_i,k_{i+1}))$ an associated
localization measure $\mA$
was computed on a sample of consecutive chaotic levels around
$k_0=\bar{k_i} = (k_i + k_{i+1})/2$,
which is a mean value of $k$ on the interval $(k_i, k_{i+1})$.
First, the separation of eigenstates, regular and chaotic,
has been done, using $M_t=0.5$, and then the chaotic eigenstates have been studied.
Moreover, the obtained distribution functions $P(A)$ were calculated
for 18 values of $\lambda$ and 9 to 12 values of $k_0$, and some selection
of them is presented and discussed in the previous section \ref{sec4}.

\begin{figure}[H]
  \centering
  \includegraphics[scale=0.70]{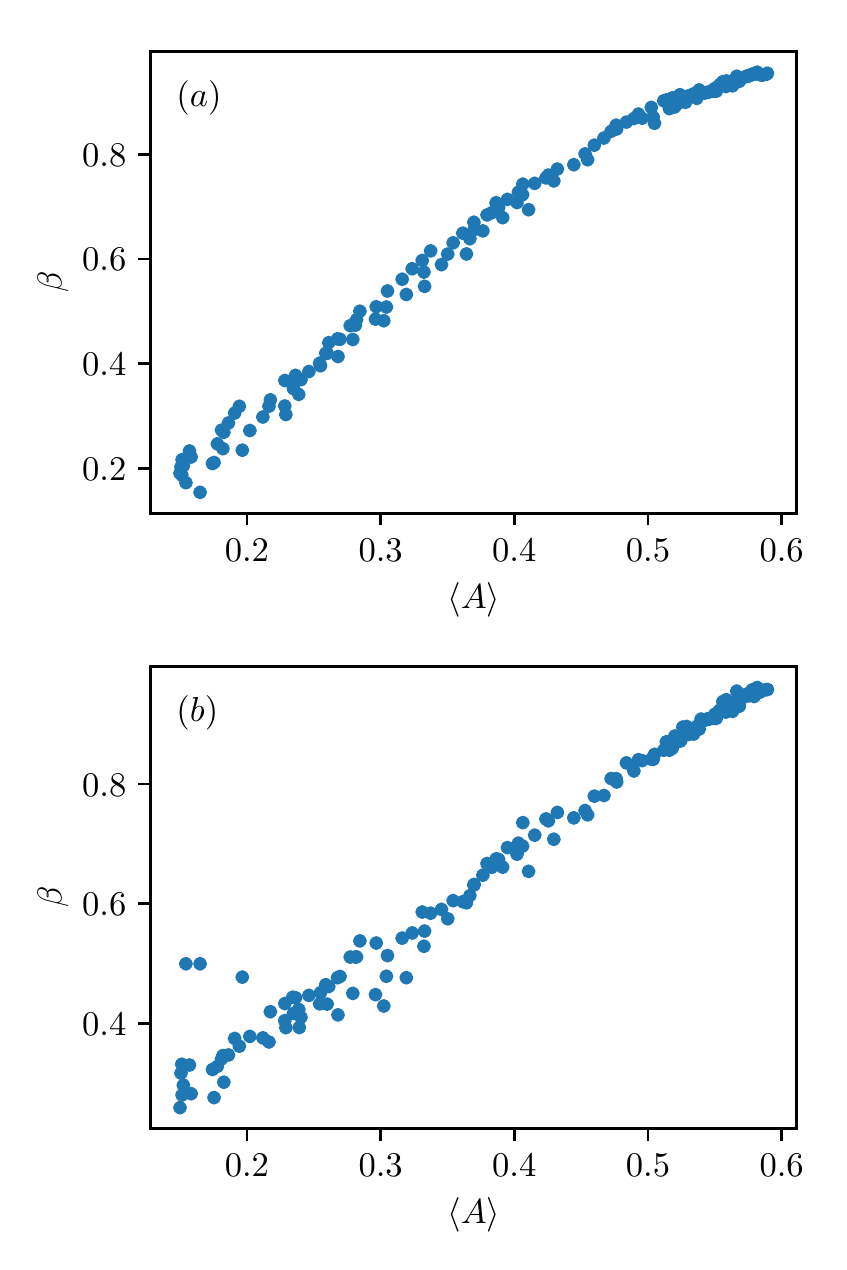}
  \caption{The level repulsion exponent $\beta$ as a function of the
    entropy localization measure $\mA$ for variety of $\lambda$
    and energies $E=k^2$, as defined in the text.  In (a) we use
    the classical criterion for $\rho_1$ and in (b) the quantum one,
    when calculating $\bet$ by fitting with BRB ditribution.}
  \label{lambdabetaVsA}
\end{figure}

The dependence of $\beta$ on $\mA$, now revised and slightly different
than  obtained in \cite{BatRob2013B}, where only one value of $\la=0.15$ was used,
is shown  in Fig. \ref{lambdabetaVsA}. We show two versions of this plot,
one with the classical criterion for $\rho_1$ (for the BRB distribution),
and the other one using the quantum criterion  for $\rho_1$,
when determining $\bet$ by fitting the level spacing distribution with the BRB
distribution. The two plots are quite similar, which is satisfactory.
This relation $\bet(\mA)$ is similar to the
case of the quantum kicked rotator
\cite{Izr1990,ManRob2013,BatManRob2013}  and the stadium. In both cases the scattering
of points around the mean linear behaviour is significant, and
it is related to the fact that the localization measure $A$ 
of eigenstates has some distribution $P(A)$, as observed and discussed in
Ref. \cite{ManRob2015} for the quantum kicked rotator, and discussed
for the stadium billiard in the Refs. \cite{BLR2018,BLR2019}.

There is still a great lack in theoretical understanding of the 
physical origin of this phenomenon,
even in the case of (the long standing research on) 
the quantum kicked rotator, 
except for the intuitive idea, that energy spectral properties should be 
only a function of the degree of localization, because the localization
gradually decouples the energy eigenstates and levels, switching the linear
level repulsion $\beta=1$ (extendedness) to a power law
level repulsion with  exponent $\beta < 1$ (localization). 
The full physical explanation is open for the future.

\begin{figure}[H]
  \centering
  \includegraphics[width=9cm]{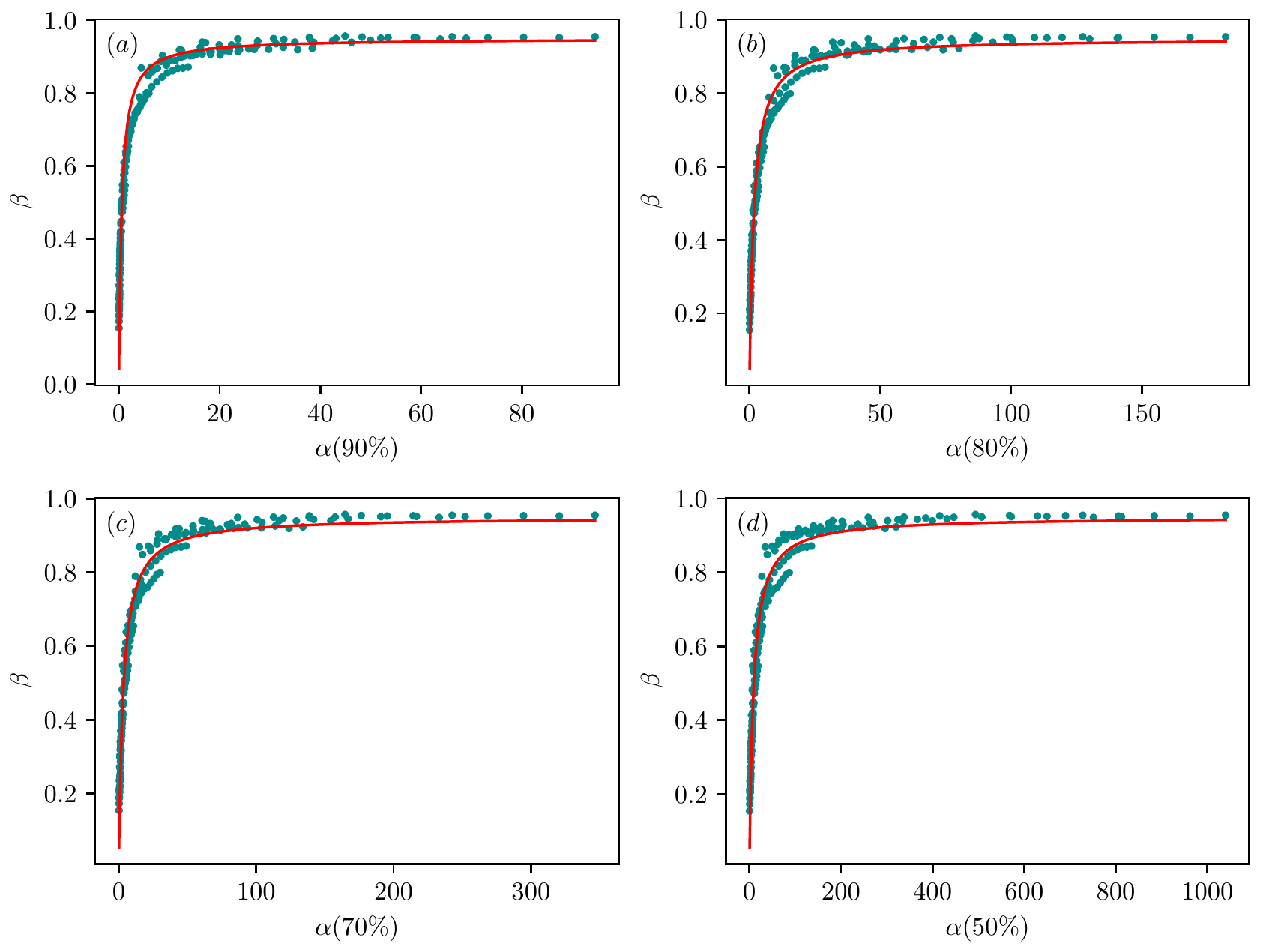}
  \caption{The level repulsion exponent $\beta$, using the classical criterion for $\rho_1$
    for the fitting BRB distribution, as a function of $\al$ fitted
    by the function (\ref{BvsAExp}), based on $N_T$ from the Table I.
    $\beta_{\infty}=0.98$ and $s=1.70,\; 0.57,\; 0.30,\; 0.11.$, for (a-d), respectively.}  
  \label{betavsalphaclass}
\end{figure}

As shown in Fig. \ref{betavsalphaclass}, using the
classical criterion for $\rho_1$ for the fitting BRB distribution,
the functional dependence of $\bet (\al)$ is always the rational function

\be  \label{BvsAExp}
\beta = \beta_{\infty} \frac{s\al}{1 +s \al}.
\ee
only the coefficient $s$ depends on the definition of $N_T$ and $\al$.
For the parameter values we get $\beta_{\infty}=0.98$ and $s=1.70,\; 0.57,\; 0.30,\; 0.11.$.
Similarly, we find for the quantum criterion almost the same results, with no visible differences
(not shown).

%

\section{Conclusions  and discussion}
\label{sec6}

In this paper we have studied the structural and statistical properties of the eigenstates and their Poincar\'e-Husimi functions, and of the energy spectra, of a mixed-type billiard \cite{Rob1983,Rob1984}, in correspondence with its classical dynamics. The governing control parameter is $\al=t_H/t_T$, where $t_H=2\pi\hbar/\Delta E$ is the Heisenberg time and $t_T$  the classical transport time, as in Eq. (5), in the semiclassical regime of sufficiently small effective Planck constant, which is $1/k$ (the wavelength).
 
Our main conclusions are as follows: {\bf (a)} We have confirmed that the normalized inverse participation ratio $nIPR$ and the information entropy measure $A$ are linearly related and thus equivalent, in agreement with the recent result in the stadium billiard \cite{BLR2019}, which we believe is a general result, not specific of the used model systems. {\bf (b)} We have calculated the Poincar\'e-Husimi functions of all eigenstates for 18 different values of the shape parameter $\lambda$ and 9 to 12 values of the starting $k_0$, in each case 2000 eigenstates of even parity. We have shown a selection of typical Poincar\'e-Husimi functions. {\bf (c)} Then we   have separated the regular and chaotic eigenstates, and verified that the chaotic states are localized to the various degree, and calculated the corresponding localization measure $A$ for all of them. {\bf (d)} We have looked at the distribution functions $P(A)$ and $W(A)$ (histograms and cumulative distributions), and found that in the regime of uniform chaos (no significant stickiness regions in the classical phase space) they are perfectly well described by the beta distribution, which in the limit of $\al \gg 1$ approaches the Dirac delta distribution $\delta(A_0-A)$. This behaviour is the same as in the stadium billiard. {\bf (e)} In the regime of existing pronounced stickiness regions in the classical phase space, $P(A)$ is not universal and can have several maxima (usually it is bimodal), where each minor maximum might be qualitatively attributed to a stickiness region. This phenomenon is under investigation in the lemon billiards \cite{LLR2019}. {\bf (f)} We have explored  the mean value $\mA$ as a function of $\al$, which is approximately a rational function, while the standard deviation of $P(A)$, denoted by $\sigma$, as a function of $\al$, exhibits strong fluctuations but nevertheless displays a similar structure as in the stadium billiard. {\bf (g)}  The level spacing distribution of localized chaotic eigenstates displays the Brody distribution, where the level repulsion exponent $\bet$ goes from $0$ for the strongest localization (Poissonian distribution) to $1$ for complete delocalization (ergodicity and GOE).  It is a function of $\mA$, but slightly different from the result in Ref. \cite{BatRob2013B}, where only one value of $\la=0.15$ has been used. It is closer to the linear relationship, which has been observed in the quantum kicked rotator and in the stadium billiard, but it also depends slightly on the criterion for choosing $\rho_1$ in determining  $\bet$.
It must be emphasized that the  transition from strong localization $\bet=0$ to  ergodicity $\bet=1$ as a function of $\al$ is a rather smooth one, not a discrete jump, as it takes place over an interval of more than factor $10$ in $\al$, which is the same behaviour as in the stadium billiard.

We believe that most of our results are quite general, typical for the mixed-type Hamiltonian systems, and where a comparison is applicable, they agree with the previous results on the stadium billiard \cite{BLR2018,BLR2019}.  Similar extensive analysis is being performed for the lemon billiard \cite{Lozej2019,LLR2019}, introduced
in Ref. \cite{HelTom1993} and studied in Refs. \cite{LMR1999,LMR2001,MHA2001,BCPV2019}, whose classical phase space has been very recently extensively explored by Lozej \cite{Lozej2019}.
The major open theoretical  question is to derive the existence of dynamical localization in chaotic eigenstates, and to calculate the corresponding  $P(A)$, and also the level repulsion exponent $\bet$  which governs the level spacing distribution, the underlying distribution being close to the Brody distribution. This problem is not yet solved even for the quantum kicked rotator \cite{Izr1990}.  Further theoretical work is in progress. We expect similar results in the smooth Hamiltonian systems of the mixed-type, such as  e.g. the hydrogen atom in strong magnetic field \cite{Rob1981,Rob1982,HRW1989,WF1989,RWHG1994} and the Dicke model  \cite{Dicke1954,FNP1998,BLLH2015}, as well as in experiments such as the microwave resonators introduced and performed by St\"ockmann since 1990 \cite{Stoe}.

\section{Acknowledgement}

This work was supported by the Slovenian Research Agency (ARRS) under
the grant J1-9112.

\providecommand{\noopsort}[1]{}\providecommand{\singleletter}[1]{#1}%

\end{document}